\documentclass[prb,amssymb,twocolumn,showpacs]{revtex4}

\usepackage{graphicx}
\usepackage{bm}

\begin{document}

\title{Quantum  chaos in the  mesoscopic device for
the Josephson flux qubit}

\author{Ezequiel N. Pozzo}
\author{Daniel Dom\'{\i}nguez}
\author{Mar\'{\i}a Jos\'{e} S\'{a}nchez}

\affiliation{Centro At{\'{o}}mico Bariloche and Instituto Balseiro,
8400 San Carlos de Bariloche,
R\'{\i}o Negro, Argentina.}

\begin{abstract}

We study the quantum spectra and eigenfunctions of the three-junction SQUID device designed for the
Josephson flux qubit at high energies. We  analyze the spectral statistics on the parameter
region where the system has a mixed classical phase space where regular and chaotic orbits can
be found at the same classical energy. We perform a numerical calculation of eigenvalues and
eigenstates for different values of the ratio of the Josephson and charging energies, $E_J/E_C$, which
is directly related to an effective $\hbar$ parameter.
We find that the nearest neighbour distributions $P(s)$ of the energy level spacings are well fitted 
by the Berry-Robnik  theory employing as free parameters the pure classical measures of the chaotic and regular regions of phase  space in the different energy regions in the semiclassical case.
The phase space representation of the wave functions is obtained via the Husimi distributions and the localization of the states on classical structures is analyzed.
We discuss for which values of $E_J/E_C$ it can be possible to perform experiments  
that could be sensitive to the structure of a mixed classical phase space.

\end{abstract}

\pacs{74.50.+r, 05.45.Mt, 85.25.Cp, 03.67.Lx}

\maketitle

\section{Introduction}

In recent years, several types of superconducting qubits have been experimentally
proposed.\cite{nakamura,qbit_mooij,vion,martinis}
These systems consist on mesoscopic Josephson devices  and  they are  promising
candidates to be used for the design of qubits for quantum
computation.\cite{nakamura,qbit_mooij,vion,martinis,revqubits,chiorescu,fqubit_recent}
Indeed, a large effort is devoted
to succeed in the coherent manipulation of their quantum states in a controlable way.
The progress made in this case allows to have nowadays Josephson
circuits with small dissipation and large decoherence
times.\cite{vion,martinis,chiorescu,fqubit_recent}
Very recently, it has been proposed that,
due to these developments,
it could also be possible to use  mesoscopic Josephson devices for the
study of the quantum signatures  of classically chaotic systems.\cite{montangero05,mingo}
In Ref.~\onlinecite{mingo} the quantum dynamics of the Device for the Josephson Flux Qubit (DJFQ)
has been studied. In particular, it has been discussed 
how the fidelity  (or Loschmidt echo)\cite{jp} of the DJFQ 
could be studied experimentally for energies corresponding to the hard chaos regime
in the classical limit.
Here, we extend the work of Ref.~\onlinecite{mingo} by analyzing 
the possibility of studying, in the DJFQ, the mixed chaos regime (i.e., 
the energy range where there
is a coexistence of chaotic and regular orbits in the classical limit). 
To this end, standard tools of analysis of ``quantum chaos'', like spectral
statistics \cite{metha,bohigas,berryro,seligman,cederbaum84,cederbaum86,prosen,makino01,robnik05} 
and phase space distributions,\cite{wigner,husimi,hus-rev,groh} will be used.

It is by now well established  that from the analysis  of the spectral properties of quantum systems
in the semiclassical regime  it is possible to obtain information about
the underlying dynamics of the  classical counterpart. \cite{metha,bohigas,berryro,seligman,cederbaum84,cederbaum86,prosen,makino01,robnik05}
The probability distribution $P(s)$ of the spacings $s$ between successive
energy levels - the nearest neighbor spacing distribution
$P(s)$-  unveils information on the  associated   classical dynamics. For integrable systems the levels are uncorrelated, and  $P(s)$ obeys a Poisson distribution. 
For completely chaotic classical motion,  $P(s)$ follows 
the prediction of the Random Matrix Theory (RMT) \cite{metha}
and  when  time reversal symmetry is preserved
$P(s)$ is closely approximated by the Wigner distribution for the Gaussian Orthogonal Ensemble (GOE),
$P(s) \sim s \exp{(-s^{2})}$.\cite{bohigas}

Generic quantum systems do not conform  to the
above  special cases, the classical phase space typically presents mixed dynamics,
with coexistence of regular orbits  and chaotic motion.\cite{berryro,seligman,cederbaum84,cederbaum86,prosen,makino01,robnik05} 
In this generic case Berry and Robnik \cite{berryro} proposed  an analytical expression for the corresponding $P(s)$, based on the knowledge  of  pure classical quantities related to  the Liouville measure of the chaotic and regular classical regions. The idea behind their calculations is
that  each regular or irregular phase space region gives rise to its own sequence of energy levels. For each region the level density results proportional to the Liouville measure of the classical region and the associated level spacing distribution follows the Poisson or the Wigner form for regular and chaotic regions respectively.
In the semiclassical limit these sequences of energy levels can be supposed independent and the complete  distribution $P(s)$ is obtained by their random superposition.
Several works have studied numerically the  level statistics in systems with mixed dynamics. \cite{seligman,cederbaum84,cederbaum86,prosen,makino01,robnik05}
Systems with two degrees of freedom
have been analyzed by several groups, mostly quartic oscillators\cite{seligman,cederbaum84,cederbaum86} 
and billiards,\cite{prosen,makino01,robnik05} 
and in some works the Berry-Robnik proposal has
been tested in detail.\cite{cederbaum86,prosen,makino01}

In contrast to the level statistics, the wave functions of quantum chaotic systems have  remained relatively less explored. In particular the analysis of wave functions in phase space representations, such as the Wigner  function \cite{wigner} or the Husimi distribution, \cite{husimi,hus-rev,groh}  allows a direct comparison between the classical and the quantum dynamics.  Of particular interest are the zeros of the Husimi distribution
 which seem to be organized along regular lines or fill space regions for regular or chaotic classical dynamics
 respectively. \cite{leboeuf}

Besides the importance of visualizing the dynamical properties of quantum systems in phase space, 
techniques for
measuring these functions, referred as ``quantum tomography " \cite{nielsen,miquel} are subjects of active research in many experimental  systems, 
like  ion traps, optical lattices, entangled photons,\cite{mitchell,kanem}
and also superconducting qubits.\cite{tomo_super}

Josephson junctions have been used for the study of classical chaos since
the early 1980s.\cite{jchaos,jchaos_exp} A single underdamped junction with
a periodic current drive can become chaotic in a wide range of parameter
values.\cite{jchaos} Several
experiments have indeed studied this problem and measured chaotic properties
in current-voltage curves and in voltage noise in
Josephson junctions.\cite{jchaos_exp}
Moreover, networks with several
junctions have been proposed for the study of spatio-temporal chaos.\cite{jchaos_arrays}
All this cases correspond to classical chaos in dissipative systems with
a time-periodic drive.
Much less studied has been the case of classical hamiltonian chaos in Josephson junctions,\cite{parmenter}
mainly due to the fact that dissipation through a shunt resistance and/or coupling
to the external measuring circuitry is typically important.
For the same reason,  {\it i.e.}, the difficulty in fabricating Josephson circuits
with negligible coupling to the environment, 
few examples of quantum chaos in Josephson systems are found in the literature. 
One  of them  is the work of Graham {\it et al.},\cite{graham}
who considered dynamical localization and level repulsion in a single Josephson junction
with a time periodic drive. 
More recently  T. D. Clark, M. J.
 Everitt and coworkers\cite{everitt}  explored chaos and the quantum behaviour 
 of SQUID rings coupled to electromagnetic field modes.
The recent development of Josephson devices for quantum computation, which need
large coherence times, lead to significant advances in the fabrication of circuits
with small coupling to the external circuit and negligible dissipation.
This opened the possibility of using this type of mesoscopic devices for the
study of quantum chaos.
For example, Montangero {\it et al.} \cite{montangero05} have proposed recently
a Josephson nanocircuit as a realization of the quantum kicked rotator. The
difficulty in realizing experimentally their system resides in that it
needs to move mechanically one superconducting node in a high-frequency
periodic motion.
A different proposal has been put forward in Ref.\onlinecite{mingo},
where it has been shown that
the Device for the Josephson Flux Qubit (DJFQ),\cite{qbit_mooij,chiorescu,fqubit_recent}
which consists on  a three-junction SQUID,
is classically chaotic at high energies.
It could be therefore possible to use this system for the experimental study of quantum signatures
of classical chaos. 
One possibility
is the analysis of the fidelity or Loschmidt echo\cite{jp} in the quantum dynamics.\cite{mingo}
An experimental setup for the measurement of the Loschmidt echo in the DJFQ has been proposed
in Ref.\onlinecite{mingo}. In the above mentioned work, 
the system is prepared initially with a wave packet\cite{nota2}
localized in coordinate (phase) and momentum (charge) with an energy corresponding to
the regime of hard chaos in the classical limit. The quantum evolution of the wave packet
is evaluated in the unperturbed and the perturbed hamiltonians, and the overlap of the
two evolved wave functions defines the Loschmidt echo or fidelity\cite{jp}, which
can be measured experimentally.\cite{mingo}  
Different behavior could be observed if the  wave packet is initially localized in 
a chaotic or in a regular region of the phase space. 
Therefore, an interesting case to analyze is 
when the wave packet is prepared initially with an energy within the regime where
there is a mixed phase space in the classical limit. In this case, one would expect
that the behavior of the Loschmidt echo could depend 
on the location of the average coordinate and momentum of the initial wave packet.
For example, in 
Ref.\onlinecite{liu} it has been found a strong dependence of the fidelity
with the initial  state for mixed dynamics in the phase space in the case 
of Bose-Einstein condensates.
However, in order to be sensitive to the structure of phase space in the 
case of mixed dynamics, it is necessary to have a small effective $\hbar$. 
The aim of the present work is to analyze the quantum spectra and wave
functions of the DJFQ  in order to obtain for which values of 
the effective $\hbar$ the quantum physics of this system can show signatures
of the structure of the phase space in the case of mixed dynamics.
To this end, we will use standard tools of quantum chaos theory
by calculating numerically the level statistics of 
the DJFQ for different effective $\hbar$
and the Husimi distribution.

Concerning the spectral analysis, 
the quantum signatures of chaos have been discussed through the $P(s)$ distribution
in Ref. \onlinecite{kato} for a SQUID with three junctions in the hard chaos regime. 
However, the case with only on-site capacitances was considered there 
(the capacitance of the junctions was neglected).
Nevertheless, the device for the Josephson flux qubit fabricated by
the Delft goup\cite{chiorescu}
has small on-site  capacitances, about two orders of magnitude
smaller than the intrinsic capacitances of the junctions.\cite{nota3}
This fact turns  the model  hamiltonian
for the DJFQ  to be different from the one studied in Refs.\onlinecite{parmenter, kato}.
One of the  goals of this paper is to  analyze the spectral properties of the DJFQ considering
realistic values of the  different capacitances to analyze 
the device for the Josephson flux qubit (DJFQ) in the case of mixed dynamics.
In addition we analyze the structure of the Husimi  functions for the DJFQ, 
an issue that has been so far unexplored.
The paper is organized as follows.
In Sec.\ref{model} we introduce the quantum  model for the device for the Josephson flux qubit.
Before presenting the quantum spectral analysis,   we will  study  in  Sec. \ref{qchaos} the dynamics of its classical analog. The presence of chaos will be characterized through the analysis of
 a measure of the chaotic volume, that  will be defined and   obtained as a function of the energy.
We devote the rest of Sec. \ref{qchaos} to the analysis of the spectral properties. 
The NNS distribution will be obtained for different energies corresponding 
to different classical energy regions and dynamics and
for different values of the effective $\hbar$.
In Sec. \ref{distri} we  compute the Husimi distribution for the DJFQ in order  
to characterize the localization of the quantum states on typical  phase space structures 
related to the different classical regimes.
Finally in Sec. \ref{conclu} we summarize our results  
and discuss possible experimental characterizations of
the quantum manifestations of chaos in this system.

\section{Model for the Device for the Josephson Flux Qubit}
\label{model}
\begin{figure}[th]
\begin{center}
\includegraphics[width=0.8\linewidth]{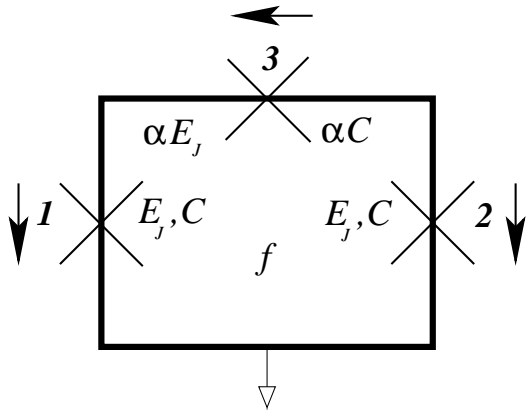}
\caption{Circuit for the Device for the Josepshon Flux Qubit as described in the text.
Josepshon junctions $1$ and $2$ have Josepshon energy $E_J$ and capacitance $C$, and junction $3$
has Josepshon energy and capacitance $\alpha$ times smaller.
The arrows indicate the sign
convention for defining the gauge invariant phase differences. The circuit encloses
a magnetic flux $\Phi = f \Phi_0$.} \label{djfq_fig}
\end{center}\end{figure}

The DJFQ consists of three
Josephson junctions in a superconducting ring \cite{qbit_mooij}
that encloses a magnetic flux $\Phi= f\Phi_0$,
with $\Phi_0=h/2e$, see Fig.\ref{djfq_fig}.

The junctions have gauge invariant phase differences defined
as $\varphi_1$, $\varphi_2$ and $\varphi_3$, respectivily, with
the sign convention corresponding to the directions
indicated by the arrows in Fig.\ref{djfq_fig}.
Typically the circuit inductance
can be neglected  and the phase difference of the
third junction is:
$\varphi_3=-\varphi _1 +\varphi _2-2\pi f$.
Therefore the system can be described
with  two dynamical variables: $\varphi_1,\varphi_2$.
The circuits that are used for the Josephson flux
qubit have
two of the junctions with the same coupling
energy, $E_{J,1}=E_{J,2}=E_J$, and capacitance, $C_1=C_2=C$,
while the third junction has smaller
coupling $E_{J,3}=\alpha E_J$ and capacitance $C_3=\alpha C$,
with $0.5<\alpha<1$.
The above considerations lead to the Hamiltonian \cite{qbit_mooij,nota}
\begin{equation}
\label{hamil}
{\cal H}=\frac{1}{2}{\vec {P}}^T
{\rm {\bf M}}^{-1}{\vec {P}}
+E_J V(\vec {\bf \varphi})\label{ham_clas}
\end{equation}
where the two-dimensional coordinate is $\vec{\varphi}=(\varphi_1,\varphi_2)$.
The potential energy is given by the Josephson energy of the circuit and,
in units of $E_J$, is:
\begin{equation}
\label{eq:pot}
V(\vec {\bf \varphi})=
2+\alpha -\cos \varphi_1-\cos \varphi_2
- \alpha \cos (2\pi f+\varphi _1 -\varphi _2 ) \; .
\end{equation}
The kinetic energy term is given by the electrostatic energy of the circuit, where
the two-dimensional momentum is
$$\vec{P} = (P_1,P_2)={\rm{\bf M}}\cdot \frac{d{\vec{\varphi}}}{dt},$$
and
$\bf M$ is an effective mass tensor determined by the capacitances of the circuit,
$$
{\rm {\bf M}}= C {\left(\frac{\Phi_0}{2\pi}\right)^2} {\rm {\bf m}}
$$
with
$$
{\rm {\bf m}}=\left(
{{\begin{array}{cc}
 {1+\alpha +\gamma }  & {-\alpha }  \\
 {-\alpha }  & {1+\alpha +\gamma }  \\
\end{array} }} \right).$$
We included in  $\bf M$ the  on-site capacitance $C_g=\gamma C$.
(Typically $\gamma \sim 10^{-2}-10^{-3}\ll 1$).
In the presence of gate charges $Q_{g,i}$ induced in the islands, the momentum
is
${\vec {P}} \rightarrow {\vec {P}} + \frac{\Phi_0}{2\pi}\vec{Q_g}$.\cite{qbit_mooij}
The system modelled with Eqs.~(\ref{hamil})-(\ref{eq:pot}) is
analogous to a particle with anysotropic mass ${\rm {\bf M}}$
in a two-dimensional periodic potential $V(\vec {\bf \varphi})$.\cite{geisel}


In typical junctions,  the Josephson energy scale, $E_J$, is much larger than
the electrostatic energy of electrons, $E_C= e^2/2C$, and the system
is in a classical regime. On the other hand, mesoscopic junctions (with small area) can have
$E_J\sim E_C$, and quantum fluctuations become important.\cite{likharev}
In this case, the quantum momentum operator  is defined as
$${\vec {P}} \rightarrow \hat{\vec{P}}= -i\hbar\nabla_\varphi = -i\hbar(\frac{\partial}{\partial\varphi_1},\frac{\partial}{\partial\varphi_2}).$$
After replacing the above defined operator $\hat{\vec{P}}$ in the
Hamiltonian of Eq.(\ref{hamil}), the  eigenvalue Schr\"odinger equation  becomes
\begin{equation}
\label{eq:Schro}
\left[ -\frac{\eta^2}{2}\nabla_\varphi^T{\rm{\bf m}}^{-1}\nabla_\varphi
+V(\vec {\bf \varphi})\right] \Psi(\vec {\bf \varphi}) = E \Psi(\vec {\bf \varphi}) \; ,
\end{equation}
where we normalized energy by $E_J$ and momentum by
$\hbar/\sqrt{8E_C/E_J}$. We see in Eq.(\ref{eq:Schro})
that the parameter $\eta=\sqrt{8E_C/E_J}$ plays the role of an effective
$\hbar$. It is well-known that the ratio $E_C/E_J$ controls
the effect of quantum fluctuations in single Josephson junctions\cite{schon,haviland}
 and in arrays of  several Josephson junctions.\cite{fazio,mooij_qjja}
For  $E_J\gg E_C$, ($\eta \ll 1$), the
junctions can be described with a classical dynamics;
while for $E_J \sim E_C$, ($\eta \sim 1$)  the effect of quantum
fluctuations becomes important.\cite{schon}
Experiments where the Josephson junctions are fabricated for different values of
$E_C/E_J$ have been performed both for single junctions\cite{haviland} 
and for junction arrays.\cite{mooij_qjja}
In the last case quantum phase transitions as a function of $E_C/E_J$
have been studied.\cite{fazio,mooij_qjja}
Therefore, the parameter $\eta=\sqrt{8E_C/E_J}$ is a natural choice
for quantifying the effective $\hbar$ in this system.

For quantum computation implementations \cite{qbit_mooij,chiorescu,fqubit_recent}
the DJFQ is operated at magnetic
fields near the half-flux quantum ($f= f_0+\delta f$, with $f_0=1/2$).
For values of $\alpha \ge 1/2$, the potential Eq.(\ref{eq:pot})
has two well defined minima.
At  the optimal operation point $f=1/2$, the two lowest (degenerated) energy states  are
symmetric and antisymmetric superpositions of two
states corresponding to macroscopic persistent currents
of opposite sign. The offset value $\delta f$ determines the level splitting
between these two states.
These  eigenstates are energetically separated
from the others (for small $\delta f$)
and therefore the DJFQ has been  used as a qubit \cite{qbit_mooij,chiorescu,fqubit_recent}
({\it i.e.} a two-level truncation of the Hilbert space is performed). 
In addition the barrier for quantum tunneling between the states depends strongly on
value of $\alpha$ and  its height goes up as  $\alpha$ is increased. 
The possibility  to manipulate the potential landscape  by changing $\alpha$
is a crucial point for experimental implementation of qubits. Typical
experiments in DJFQ have values of $\alpha$ in the range $0.6-0.8$.\cite{chiorescu,fqubit_recent}

As we will discuss here, the higher energy states
of the DJFQ show quantum manifestations of classical chaos.
In what follows we focus our study of the DJFQ considering the
realistic case of: (i) small on-site  capacitances,
taking $\gamma=0.02$, (ii) a magnetic field corresponding to the optimal operation
point of the DJFQ, $f=1/2$, and  (iii) the values of $\alpha= 0.7$ and $0.8 $  
in coincidence with the experimental  values employed  in
Ref. \onlinecite{chiorescu,fqubit_recent}. 


\section{Spectral Statistics}
\label{qchaos}

Before entering into the analysis of the  quantum spectra we will
focus on  the classical dynamics of the DJFQ.
As we  already anticipated in the Introduction, generic systems present
mixed classical dynamics and the DJFQ is not the exception.
Therefore for a given energy $E$ our aim is to  estimate the chaotic volume $v_{\rm ch}(E)$,
defined as the probability of having a chaotic orbit (i.e. Lyapunov exponent $\lambda > 0$) at energy $E$.
As we will show below, this parameter  will be relevant  in  the statistical  analysis of the quantum spectrum.

The classical dynamical evolution was obtained solving the Hamilton equations derived from Eq.(\ref{hamil}):
\begin{equation}
{\rm {\bf m}}\cdot \frac{d^2{\vec{\varphi}}}{dt^2} = -\nabla_\varphi V(\vec {\bf \varphi}),
\end{equation}
where we have normalized energy by $E_J$ and time by $t_c=\sqrt{\hbar^2C/4e^2E_J}
=\hbar/\eta E_J$ (the Josephson plasma frequency is $\omega_p=t_c^{-1}$).
The numerical integration was performed with a second-order leap-frog algorithm
with time step $\Delta t = 0.02 t_c$.

\begin{figure}[th]
\begin{center}
\includegraphics[width=0.9\linewidth]{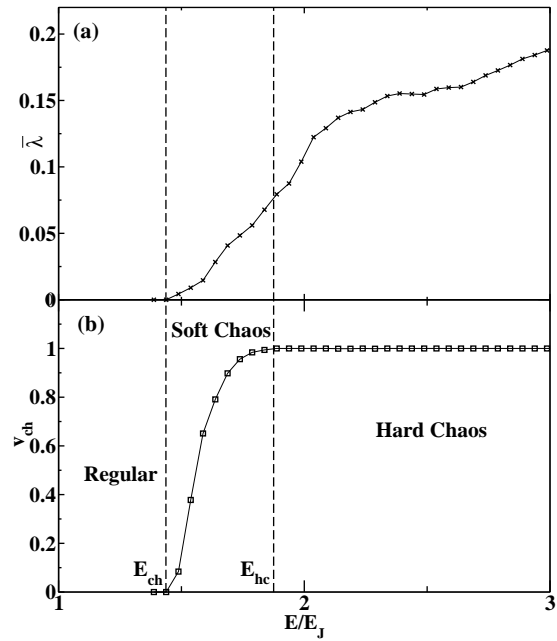}
\caption{ (a) Average maximum Lyapunov
exponent $\overline{\lambda} $ and (b) chaotic volume $v_{ch}$
 versus energy $E$ for $\alpha=0.8$
and $f=1/2$. } \label{figchaos}
\end{center}\end{figure}
\noindent

For different values of the parameter $\alpha$ and
magnetic flux $ f $ we compute the maximum Lyapunov exponent
$\lambda$ for each classical orbit at different energies $E$. We estimate
the chaotic volume $v_{\rm ch}(E)$ using $10^3$ initial conditions chosen randomly
with uniform probability within the available phase space for each
given energy. Also the average Lyapunov exponent, $\bar{\lambda}(E)$, of
the chaotic orbits is obtained.
These results are shown in Fig.~\ref{figchaos} for $\alpha=0.8$ and $f=1/2$.
We observe that both $v_{\rm ch}(E)$ and  $\bar{\lambda}(E)$
increase smoothly with energy, as it is usual in several similar
systems with two degrees of freedom.\cite{benettin,meyer,cejnar}
Above the minimum energy of the potential, $E_{min}$, we find: (i) {\it
regular orbits} for $E_{min} < E < E_{ch}$ ($v_{\rm ch}=0$), (ii) {\it
soft chaos} ({\it i.e.}, coexistence of regular and chaotic
orbits, $0 < v_{\rm ch} < 1$) for $E_{ch} < E < E_{hc}$ with
the average Lyapunov exponent $\bar{\lambda}>0$ above $E_{ch}$
 and (iii) {\it hard chaos} (all orbits are chaotic, $v_{\rm ch}=1$) for
$E>E_{ch}$.
The boundaries  of these different dynamic regimes
as a function of $\alpha$, in the range $[0.5,1.0]$, and $f$, in the
range $[0,0.5]$, have been obtained in Ref.~\onlinecite{mingo}.
Here we will focus on the case with $f=1/2$ and we will study  some different
cases of $\alpha$.

In order to look for signatures of quantum chaos, we follow a standard
statistical analysis of the energy spectrum.
First we calculate the exact spectrum $\{E_n\}$ by diagonalization of the
quantum Hamiltonian. The eigenvalue equation Eq.(\ref{eq:Schro}) is solved
by discretizing the phases with $\Delta \varphi = 2\pi/1000$, and the resulting
hamiltonian matrices of size $10^6 \times 10^6$ are diagonalized using
standard algorithms for sparse matrices. We have verified that
increasing the discretization by a factor of $2$ does not affect the results of the
spectrum within the needed accuracy for the ranges of energies studied here.
As we mentioned we set  $\gamma=0.02$ and   $f=1/2$,
and we obtain eigenvalue spectra for different values of the parameters
$\eta$ and $\alpha$  defined in the previous section.

The level spectrum is used to obtain the smoothed counting function
$N_{av}(E)$ which gives the cumulative number of states below an
energy $E$.
In order to analize the structure of the level fluctuations
properties one unfolds the spectrum by applying the well
kwown transformation $x_n=N_{av}(E_n)$.\cite{bohigas} From
the unfolded spectrum one can calculate the nearest-neighbor spacing (NNS)
distribution $P(s)$, where $s_i \equiv x_{i+1}-x_i$ is the NNS.

We have taken into account the symmetries of
the Hamiltonian Eq.(\ref{hamil}).
For $f=1/2$ the Hamiltonian has reflection symmetry against
the axis $\varphi_2=\varphi_1$ and against the axis $\varphi_2=-\varphi_1$.
The eigenstates can be chosen with a given parity with
respect to these two symmetry lines.  Therefore, we compute the  NNS distribution
 employing eigenstates of  a given parity.
This kind of decomposition is a standard procedure followed in the
analysis of spectral properties of quantum systems whenever the
Hamiltonian of the system possesses a discrete symmetry.\cite{bohigas}
We consider the even-even parity states and
the NNS distribution is computed for   different energy regions
inside the  classical  interval  ($E_{ch},E_{hc}$), corresponding to
soft chaos, and for energies $ E > E_{hc}$ ( and $E< 2\Delta$), corresponding
to hard chaos.

The  Berry- Robnik theory seems to be suitable to analyze, in the semiclassical regime, sequence of levels
of  quantum systems whose classical analogous presents coexistency of regular and chaotic dynamics
({\it i.e.}, soft chaos regime).
If $\rho_{1}$ and $\rho_{2} $ are the relative measures of the regular and chaotic parts of the classical phase
space then,  the Berry-Robnik  distribution \cite{berryro} reads:
\begin{eqnarray}
\label{pbr}
P^{BR} (s)  = \rho_{1}^{2} \exp{(- \rho_{1} s)} \; \mbox{erfc} \left( \frac{1}{2} \sqrt{\pi} \rho_{2} s \right) + \nonumber \\
\left( 2 \rho_{1} \rho_{2} +  \frac{1}{2} \pi \rho_{2}^{3}  s \right)  \exp \left( - \rho_{1} s -
\frac{1}{4} \pi \rho_{2}^{2} s^{2} \right) \; ,
\end{eqnarray}
where $\rho_{1} + \rho_{2} =1$. It is easy to verify that $P^{BR} (s)$ interpolates between the Poisson and Wigner GOE distributions as  $0 \rightarrow \rho_{1}  \rightarrow 1$, but does not exhibit level repulsion for $\rho_{1} \neq 0$.

\begin{figure}[th]
\begin{center}
\includegraphics[width=1.\linewidth]{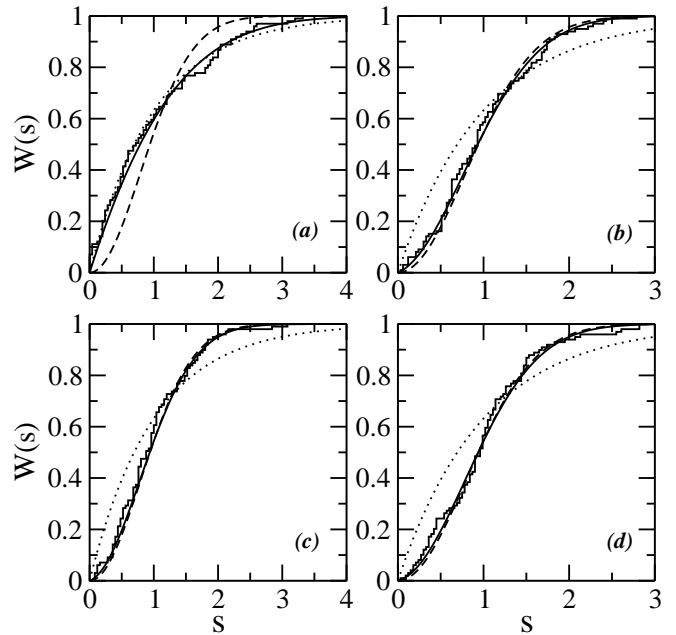}
\caption{Cumulative distribution $W(s)$ for  $\alpha=0.8$ and $f=0.5$. See the text for details.
 The continuous line is the fitted Berry-Robnik distribution. We show for comparison
 the Poisson (dotted line) and Wigner (dashed line) cumulative distributions.
 Top panels correspond to $E=1.6$ with (a) $\eta=0.01$ and (b) $\eta=0.05$.
 Bottom panels  correspond to $E=2.0$ with (c) $\eta=0.01$ and (d) $\eta=0.05$.
 The fitted Berry-Robnik parameters are  (a) $\rho_{br}=0.44$,
 (b)  $\rho_{br}=0.93$,
 (c) $\rho_{br}=0.99$ and (d) $\rho_{br}=0.96$.}
\label{nns_fig}
\end{center}
\end{figure}

In Fig.\ref{nns_fig} we show the cumulative level spacing distribution $W(s)= \int  P(s) ds$ obtained numerically
following the prescription described before. We have done this in order to describe in some detail the behavior
for small values of $s$, (in the following we denote the cumulative distributions by the same name that the
corresponding NNS distribution).
In all the cases  we have fitted the numerically obtained $W(s)$ employing
Eq.(\ref{pbr}) for the NNS distribution,
and we have   extracted the fitted  quantum parameter  $\rho_{1} \equiv \rho_{br}$.

The particular results presented in Fig.~\ref{nns_fig} correspond to a window of
$\sim 100$ eigenvalues around $E_{ch} < E=1.6 < E_{hc}$,
within the soft chaos regime, Fig.~\ref{nns_fig} (a),(b);
and $E_{hc} < E=2$, within the hard chaos regime, Fig.~\ref{nns_fig} (c),(d).
We take the realistic experimental value for the  parameter  $\alpha=0.8$ and consider different values
of the quantum parameter  $\eta$ : the case with
$\eta=0.01$ is shown in  Fig.~\ref{nns_fig} (a),(c); and the case with
$\eta=0.05$ is shown in Fig.~\ref{nns_fig} (b),(d).
We should   remark that the classical dynamics is independent of the
parameter $\eta$, which  has a pure quantum origin
and plays the role of an effective Planck's constant in
the Schr\"odinger equation, as we mentioned before.
In addition  in Fig.\ref{nns_fig} we  show for comparison the $W(s)$ corresponding to
the  Poisson and Wigner GOE distributions.

We first discuss the case with $\eta=0.05$, that  is
already smaller than in the cases studied in Ref.\onlinecite{mingo}, where
$\eta=0.07-0.17$ was considered.
The numerical results for $\eta=0.05$,  in  the case of hard chaos [$E=2.0$, 
shown in Fig.~\ref{nns_fig} (d)], are 
in good agreement with the Wigner distribution, and we obtain $\rho_{br}=0.96$. 
In a case corresponding to mixed  classical dynamics
[$E=1.6$, shown in Fig.~\ref{nns_fig} (b)], we find that
the distribution departs slightly from the pure Wigner form.
However,  we have obtained
$\rho_{br}=0.93 \gg v_{ch} \sim 0.4$, meaning that 
the level distribution in this case does not seem to be very 
sensitive to the mixed phase space expected in the classical limit.
The reason is that for increasing $\eta$ 
the mean energy level spacing increases (proportional to $\eta^2$ for
large energies), and therefore the width of the energy region evaluated for the
statistics with a given number of levels ($\sim 100$ in this case) also increases
in the same way. Since $v_{ch}(E)$ varies rapidly with $E$ within the soft chaos
region, relating its value with the fitted $\rho_{br}$, which is obtained
evaluating the statistics over a wide energy region, becomes meaningless for large
$\eta$.
Indeed, deep in the quantum regime the Berry-Robnik fitted parameters are not expected
to be related to the
classical measure of the chaotic (regular) part of the phase space.\cite{berryro,cederbaum86,prosen}

\begin{figure}[th]
\begin{center}
\includegraphics[width=0.8\linewidth]{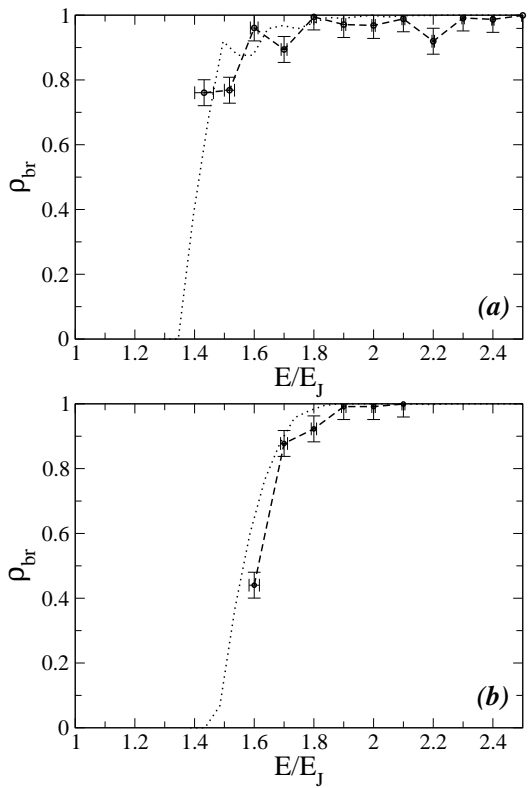}
\caption{Fitted Berry-Robnik parameter $\rho_{br}$ as a function of the
dimensionless energy $E/E_{J}$ for $f=0.5$ and $\eta=0.01$. (a) $\alpha=0.7$ and  (b) $\alpha=0.8$.
The horizontal error bars in energy are defined
by the interval of eigenenergies used in the statistics,
and  it is a decreasing function of the density of states.
The vertical error bars correspond to error in the parameter $\rho_{br}$ as
obtained from the numerical fits.
The dotted line shows the chaotic fraction of the classical phase space $v_{ch}$ obtained
from the classical dynamics.} \label{fig_br}
\end{center}
\end{figure}

We now discuss a smaller value of the effective $\hbar$,
corresponding to $\eta=0.01$.
In Fig.~ \ref{nns_fig} (a), for $E=1.6$ (mixed classical dynamics),  we find now
 that the $W(s)$ clearly departs from the pure Wigner form, and that it can be fitted with
the Berry- Robnik distribution with  $\rho_{br}=0.44$. 
This value is very close to the classical chaotic volume for this case, $v_{ch} \approx 0.4$. 
In the case for $E=2$ (hard chaos),
shown in Fig.\ref{nns_fig} (c),
we have obtained $\rho_{br}=0.99$, in agreement with $v_{ch}=1$ and also
in good agreement with the Wigner distribution, as expected.\cite{bohigas,seligman}

In general we  find that in a nearly semiclassical regime, $\eta=0.01$, the numerical results
for the Berry-Robnik parameter $\rho_{br}$ show a good agreement
with the classical measure $\rho_{1}$,
that by definition is equivalent to the chaotic volume $v_{ch}$.
This is analyzed in  Fig. \ref{fig_br} where we plot the quantum parameter $\rho_{br}$
obtained for different sections  of the spectra with $\sim 100$ eigenvalues around a given energy $E$.
We show results for two cases of the parameter $\alpha$.
The chaotic fraction of the classical phase space $v_{ch}$ is also plotted.
The results for $\rho_{br}$ and $v_{ch}$ are very close to each other.
When changing the parameter $\alpha$ the location in energy of the onsets
of the soft chaos and hard chaos regimes shifts. We also see that the
curves of $\rho_{br}$ vs. $E$ shift in the same way, giving further support
to the correspondence between $\rho_{br}$ and $v_{ch}$.\cite{seligman,cederbaum86,prosen,makino01}
 These results corroborate the validity of  the Berry-Robnik theory in the
semiclassical energy region
that corresponds to  small  effective Planck's constant, as it is the case  for $\eta=0.01$.

Besides the cases reported above, 
we have also analyzed a few other values of $\alpha$ in the range 
$0.5-0.9$ and $f=0.4,0.5$,
obtaining similar results for the spectral statistics.
In general, we observe that in order to obtain a spectral statistics with a Berry-Robnik
paremeter that agrees with the classical measure of the chaotic volume values
of $\eta < 0.05$ are needed.

\section{Phase space and Husimi Distributions for the DJFQ}\label{distri}

In this section we pursue our study of the signatures of quantum chaos 
presenting an analysis of the quantum phase-space distributions in the
case of mixed classical dynamics.
Taking into account the analysis of the previous section we focus
on the case $\eta=0.01$.

Quantum phase space distributions are of increasing interest in
studies of quantum chaos because they allow a direct comparison
between classical and quantum dynamics. The Husimi distribution
associated to a quantum wave function  $|\Psi \rangle$ (see
definition below, Eq.(~\ref{husi})) it  is based on the
coherent-state representation and is well suited to represent wave
functions in phase space because  it is always real and
possitive.\cite{husimi,leboeuf,hus-rev,groh} Due to these
properties  it is usually referred as a quasi probability
distribution.

In order to compute the Husimi function for the DJFQ we must take
into account the fact that the classical phase space is four
dimensional. The Husimi distribution function  for a state
$|\Psi\rangle$ is
\begin{equation} \label{husi}
\rho^H(\vec{P_0},\vec{\varphi_{0}})  = |\langle \vec{P_0},\vec{\varphi_{0}}|\Psi\rangle|^2\;,
\end{equation}
where
$| \vec{P_0},\vec{\varphi_{0}}\rangle$  corresponds to minimum-uncertainty
$2\pi$-periodical wave packets \cite{carruthers} given by
\begin{eqnarray}\label{eq:packet}
|\vec{P_0},\vec{\varphi_{0}}\rangle &=& C\times
\exp[i \vec{K_0}\cdot \left(\vec{\varphi}-\vec{\varphi_0}\right)] \times
\\
& &
\exp\left[\frac{\cos(\varphi_{0,1}-\varphi_1)+\cos(\varphi_{0,2}-\varphi_2)-2}{2\sigma^2}\right]
\nonumber
\end{eqnarray}
where $\vec{K_0}=(k_1,k_2)$ with $k_1,k_2$ integers
and $\vec{P_0}=\eta \vec{K_0}$.
The width of the  wave packet is
$\sigma=\sqrt{\eta/s}$,  with $s$ the squeezing parameter,
and we choose  the value $s=3.23$, which is the
same value used in Ref.\onlinecite{mingo} for the
initial coherent wave packets.

The potential has two minima for $f=1/2$ which are at $(\varphi^*,-\varphi^*)$
and $(-\varphi^*,\varphi^*)$, with $\cos\varphi^* = 1/2\alpha$.  To better
analyze the Husimi function, it is convenient to make the following change of variables:
\begin{eqnarray}\label{eq:rotation}
\varphi_x &=& \frac{\varphi_1-\varphi_2}{\sqrt{2}},\;\;\;\;\;\;\;
P_x = \frac{P_1-P_2}{\sqrt{2}};\nonumber\\
\varphi_y &=& \frac{\varphi_1+\varphi_2}{\sqrt{2}},\;\;\;\;\;\;\;
P_y = \frac{P_1+P_2}{\sqrt{2}}.
\end{eqnarray}
In this way the two minima lie along the direction of $\varphi_x$.
The normalization by $\sqrt{2}$ is chosen such that
new variables satisfy $[\varphi_x,P_x]=i\eta$, $[\varphi_y,P_y]=i\eta$
in the quantum regime.

The classical Poincar\'e surface of section is calculated in the
plane $(\varphi_x, P_x)$, taking $\varphi_y = 0$ and $P_y>0$.
We want to compare the Husimi distribution $\rho^H_\nu(\vec{K},\vec{\varphi})$ corresponding to
the eigenstate $|\Psi_\nu\rangle$ with eigenvalue $E_\nu$ with the classical
Poincar\'e section at an energy $E\approx E_\nu$.
To this end, we construct an  analog of the surface of section
by obtaining a two-dimensional section of  $\rho^H_\nu(\vec{K},\vec{\varphi})$
(which is a four-dimensional density in phase space) in the following way:\cite{groh}
\begin{equation}\label{eq:sec_husi}
\Phi^H_\nu(P_x,\varphi_x) = \rho_\nu^H(P_x,P_y^E;\varphi_x,0)
\end{equation}
where, given the values $P_x, \varphi_x$ and $\varphi_y=0$,
$P_y^E$ is obtained such that the classical energy is equal to $E$ and
the possitive root, $P_y^E>0$, is chosen.

We obtain numerically the eigenstates $|\Psi_\nu\rangle$ of Eq.~(\ref{eq:Schro}),
after using a discretization of $\Delta\varphi= 2\pi/500$. Then, using Eqs.~(\ref{husi})-(\ref{eq:sec_husi}),
we compute the sections of the Husimi distributions, $\Phi^H_\nu(P_x,\varphi_x)$.
In order to characterize the localization of  the  quantum states on the classical phase space structures,
we choose a few examples of $\Phi^H_\nu$ for eigenstates  that lie in energy regions
corresponding to regular classical dynamics $E<E_{ch}$ and
soft chaos region, $E_{ch}<E<E_{hc}$, respectively.

\begin{figure}[th]
\begin{center}
\includegraphics[width=0.8\linewidth]{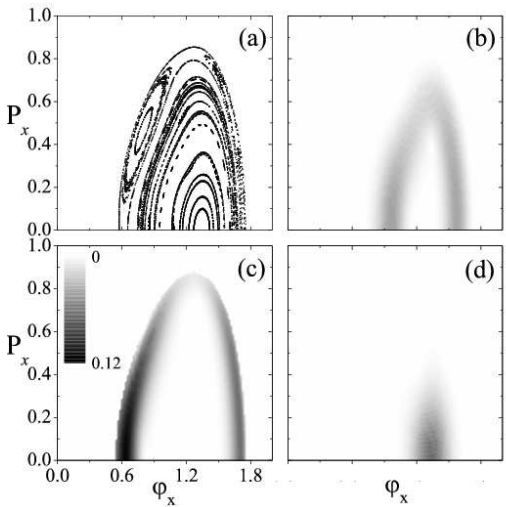}
\caption{(a) Classical Poincar\'e surface of section
for $E=1.52$. Sections are  symmetric with respect to
$\varphi_x \rightarrow -\varphi_x$
and $P_x \rightarrow -P_x$;
 only the region of $\varphi_x>0$ and $P_x > 0$  is shown.
Section of Husimi phase space distribution,
$\Phi^H_\nu(P_x,\varphi_x)$ for
(b) $E_\nu=1.5219$,
(c) $E_\nu=1.5208$,
(d) $E_\nu=1.5193$.
} \label{fig_hus1}
\end{center}
\end{figure}

In Fig.~\ref{fig_hus1} (a) we plot for $E=1.52 < E_{ch}$ the classical Poincar\'e section in which the stability
islands associated to the  regular dynamics are observed.
 We have computed the Husimi phase space distributions $\Phi^H_\nu(P_x,\varphi_x)$ 
 for several eigenstates ($\sim 20$) near the energy $E=1.52$.
 We show here three cases corresponding to eigenstates with energies $E_\nu=1.5219$,
$E_\nu=1.5208$ and  $E_\nu=1.5193$ (panels (a) , (b) and (c) respectively).
The localization of these states on the  stability islands and fixed points is clearly observed.

\begin{figure}[th]
\begin{center}
\includegraphics[width=0.8\linewidth]{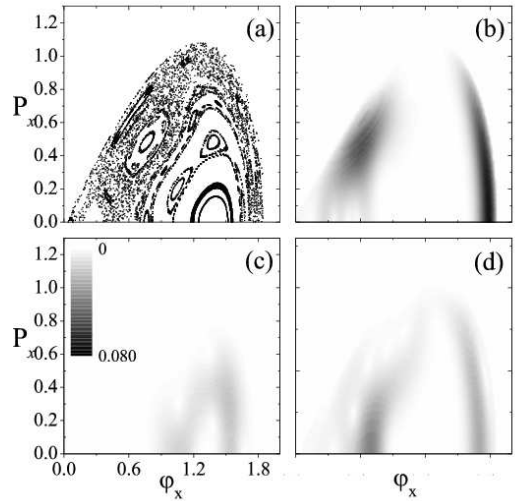}
\caption{Classical Poincar\'e surface of section
for $E=1.6$.
Section of Husimi phase space distribution,
$\Phi^H_\nu(P_x,\varphi_x)$ for
(b) $E_\nu=1.601$,
(c) $E_\nu=1.6008$,
(d) $E_\nu=1.5993$.} \label{fig_hus2}
\end{center}
\end{figure}

\begin{figure}[th]
\begin{center}
\includegraphics[width=0.8\linewidth]{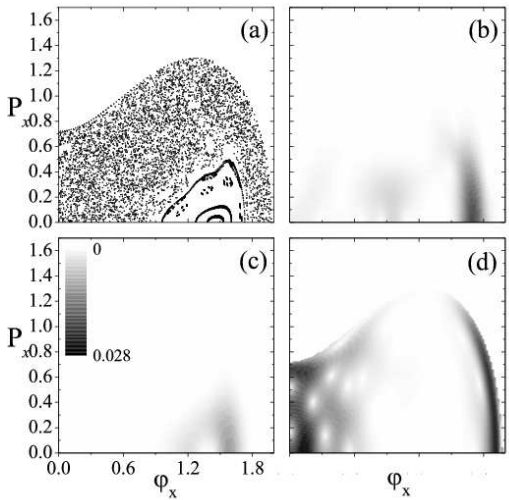}
\caption{Classical Poincar\'e surface of section
for $E=1.7$.
Section of Husimi phase space distribution,
$\Phi^H_\nu(P_x,\varphi_x)$ for
(b) $E_\nu=1.6992$,
(c) $E_\nu=1.7004$,
(d) $E_\nu=1.6994$.} \label{fig_hus3}
\end{center}
\end{figure}

In Fig.\ref{fig_hus2}(a) and Fig.\ref{fig_hus3}(a) we plot for  classical energies $E=1.6$ and $E=1.7$
respectively, the classical Poincar\'e sections  together with a selection of
some of the calculated
Husimi phase space distributions  $\Phi^H_\nu(P_x,\varphi_x)$ for eigenstates with energies
$E_\nu=1.601,1.6008, 1.5993$ (Fig.\ref{fig_hus2} (b), (c) and (d) respectively) and
$E_\nu=1.6992, 1.7004, 1.6994$ (Fig.\ref{fig_hus3} (b), (c) and (d) respectively). 

In these cases the soft chaos behavior is evident by the structure of the Poincar\'e sections,
in which regular islands are sorrounded by  chaotic regions.
The localization of the states on classical structures like  already distroyed
chains of islands  is observed in the figures. In addition, the Husimi distribution
of  Fig.\ref{fig_hus3} (d) corresponds to a
state localized on the  chaotic region of Fig.\ref{fig_hus3}(a).

The above analysis of the Husimi distributions shows that for $\eta=0.01$,
it is possible to use  localized wave packets as initial conditions for the experimental measurement of
the Loschmidt echo,\cite{mingo,nota2} 
since they can sense the structure of the phase space
with mixed classical dynamics in this case.\cite{liu}

\section{Conclusions}\label{conclu}

In this paper we have  characterized the quantum signatures of chaos in the three-junction SQUID device.
For realistic parameter values the classical dynamics exhibits different regimes that go from mixed dynamics to fully developed chaotic motion. As a consequence the spectral statistics, characterized by the distribution of the  nearest neigbour  energy spacing (NNS)  $P(s)$ in  the high
energy region, is expected to unveil  signatures of the mentioned behavior.
The analysis has been performed for different energy regions inside the classical  intervals
corresponding both to the soft chaos ({\it i.e.}, mixed phase space) and hard chaos regimes,
and we considered the even-even parity states to compute the NNS distribution.
Our numerical results show that, for $\eta < 0.05$ (and for $\eta=0.01$ in particular),
in a nearly semiclassical regime, $P(s)$ is  well fitted by
Berry-Robnik like formulae, where the pure classical measures of the chaotic and regular regions
have been used as the only free parameters.

We also found that the individual eigenstates can also be intimately linked to the phase space structures 
that characterizes the different classical regimes for $\eta<0.05$.
In order to analyze how quantum states are supported or localized on different classical structures that are present in the different regimes in this case, 
we have investigated  the Husimi  phase space distributions 
for different eigenstates with energies $E_\nu$ in the classical interval.
We would like to mention that there are few studies
of Husimi  distributions for  Hamiltonian systems with two degrees of freedom, \cite{groh}
as it is the case of the DFJQ studied in the present work.

One important advantage of  Josephson junction devices  is that they can be fabricated
with well-controlled parameters.
The effective $\hbar$, is
$\hbar_{\rm eff}=\eta=\sqrt{\frac{8E_C}{E_J}}$, and since $E_J \propto A$ and $E_C \propto 1/A$,
with $A$ the area of the junctions, we have that  $\hbar_{\rm eff}\propto 1/A$.
Thus, the fabrication of different DJFQ with junctions with varying area could allow to study
cases with $\hbar_{\rm eff} $ spanning from the semiclassical to the quantum regime.
This is indeed important since different regimes can be accessed experimentally
depending on the magnitude of $\eta$.
The qubit regime of two-level dynamics of the DJFQ is observed experimentally in
devices with $\eta \approx 0.4$.\cite{chiorescu,fqubit_recent} 
In Ref.\onlinecite{mingo} it has been found that signatures of chaos in the Loschmidt
echo can be observed at high energies $E \sim 3 E_J$ in devices
with an effective $\hbar$ of the order of $\eta \approx 0.1$.
Here we have shown that the observation of the quantum effects 
in the case with mixed chaotic and regular orbits (for an intermediate energy range) 
needs the study of devices in a more semiclassical regime
with $\eta\approx 0.01$. This could motivate experimental measurements
looking for the dependence of the Loschmidt echo\cite{mingo} with initial conditions,
due to the phase space structure of the mixed classical dynamics, if
the experiments are performed in devices with  $\eta\approx 0.01$.
Considering the  values\cite{nota} of $E_J \sim 250 GHz \sim 2 K$  
and the operation temperature of $20$ m$K$ reported by the Delft group \cite{chiorescu}, 
typical level spacings of $ 0.01 E_J  \sim 20$ m$K $ can be experimentally resolved
in the device of Ref.\onlinecite{chiorescu}.
This energy resolution is enough for the case of the Loschmidt echo in devices
with $\eta \sim 0.1$, analyzed originally in Ref.\onlinecite{mingo}.  
However, the semiclassical regime explored in this work ($\eta=0.01$) 
requires a resolution in the level spacings  of the order of $5\times10^{-4} E_J$. 
Thus, for experiments in the cryogenic range ($20$ m$K$) devices with larger values 
of $E_J$ should be  employed. 
On one hand, a smaller $\eta \sim 0.01$ already requires  junctions 
with larger area $A$, and therefore larger $E_J$.
On the other hand, Josephson junctions fabricated with high $T_c$ superconductors\cite{jhtc}
can have a large $E_J$. 
Therefore, devices designed with high $T_c$ superconductors
can be  good   candidates for the experimental  challenge of studying
the mixed phase space in the semiclassical regime of the DJFQ.

Another possible type of experiment is to start the system in the ground state and apply a constant pulse
in some external parameter (for instance, the magnetic field).
After the pulse is applied, the probability of remaining
in the ground state could be related to the energy level statistics.\cite{cohen}
Also, an interesting experiment could be to perform studies of the
low frequency noise, as it has been done in mesoscopic chaotic cavities.\cite{buttiker,beenakker-rmp}
For example, one could drive the DJFQ into the hard chaos regime with a voltage source
such that $E_V= \frac{1}{2} C V^2 > E_{hc}$ (and  $ V < 2\Delta /e$)   
and then measure the  noise in the current.
How the current noise is related to the spectral
statistics in this case is a very interesting problem, 
which could be the subject of future studies.


\acknowledgments

We acknowledge financial support
from ANPCyT (PICT2003-13829, PICT2003-13511 and PICT2003-11609),
Fundaci\'{o}n Antorchas,  CNEA  and Conicet.
ENP also acknowledges support from U.N. Cuyo.


\begin{thebibliography}{}
\bibitem{nota2}
Minimum uncertainty wave packets (i.e., coherent
states) are typically used for the numerical computation of the
Loschmidt echo as well as for analytical calculations in the semiclassical
 limit. However, ``idealized coherent states" are not
needed in a strict sense, and localized wave packets are enough
as initial conditions for the experimental observation 
of the Lyapunov regime of the Loschmidt echo.
In  Ref.\onlinecite{mingo} it is discussed how localized wave packets
could be prepared experimentally  in the DJFQ.
\end{thebibliography}
\end{document}